\documentclass[a4paper,aps,twocolumn,nofootinbib]{revtex4}
\RequirePackage[english]{babel}
\RequirePackage[latin1]{inputenc}
\RequirePackage[T1]{fontenc}
\RequirePackage{mathrsfs}
\RequirePackage{amsmath}
\RequirePackage{amssymb}
\RequirePackage{amsbsy}
\RequirePackage{color}
\RequirePackage{bm}
\RequirePackage{slashed}
\begin{document}
\title{\bf{Non-Causal Propagation for Higher-Order Interactions of Torsion with Spinor Fields}}
\author{Luca Fabbri}
\affiliation{DIME, Universit\`{a} di Genova, P.Kennedy Pad.D, 16129 Genova, ITALY}
\date{\today}
\begin{abstract}
We consider field equations of spinors with torsional interactions having higher-order dimension: by applying the Velo-Zwanziger method, we obtain that it is always possible to find situations where the propagation is affected by non-causal behaviour.
\end{abstract}
\maketitle
\section{INTRODUCTION}
Of all fields, spinors, and specifically Dirac spinors, are the foundations of the standard model of particle physics.

Dirac spinors have, beyond several gauge charges, and energy, also spin, and so they must couple, beside to those gauge fields, and to curvature, also to torsion (for general references about torsion see \cite{C1,C2,C3,S,K,h-h-k-n,Hehl:2007bn}, for the specific case of torsion coupling to the Dirac spinor see \cite{A-L,M-L,Fabbri:2006xq, Fabbri:2009se,Fabbri:2009yc, Fabbri:2008rq, Fabbri:2010rw}).

The spin-torsion interaction, despite being well known, has always been disregarded, due to a couple of misconceptions: the first is that torsion effects have always been thought as tiny, the second is that torsion-spin couplings have always been assumed equal for every spinor.

Neither assumption is true. Torsion is an independent field, specifically it is independent from curvature, and so it does not need to enter in the Lagrangian only through curvature terms, but it may enter in the Lagrangian also as square-torsion terms \cite{Fabbri:2012yg}, so shifting the overall value of the torsion-spin coupling; additionally, the torsion-spin couplings for two given spinors do not need to be equal, but they may be different \cite{Fabbri:2014naa}, therefore giving a further shift to the torsion-spin coupling for every spinor.

These two are the simplest generalizations of the spin-torsion coupling because they involve only a shift of their constants, and not the structure of the coupling between torsion and the spinor. But the spinor-torsion interaction can also be modified in its structure, if in the interaction terms we allow derivatives to take place, thus giving rise to the so-called higher-order dimensional interactions.

Higher-order dimensional interactions have been studied in \cite{Fabbri:2014dxa}, where, by employing arguments of consistency about the torsionless limit and the Velo-Zwanziger analysis, it has been possible to prove a number of interesting features: parity-evenness is required within the geometric sector; although torsion can generally be decomposed in three irreducible parts, nevertheless only the completely antisymmetric dual of an axial vector is to be kept; there can be no torsion-curvature coupling; in the most general of the cases, torsion is a propagating field of Proca type.

However, in the above work we have not deepened the higher-order dimensional interactions of torsion involving the Dirac spinor like those containing derivatives in the interaction term. As fair as it may be to stress that these interaction terms are not renormalizable, nonetheless it is still interesting to study them, both because the criterion of renormalizability has lost some of the popularity it had twenty years ago and because such interactions recently gathered some attention in view of their interest for the problem of Lorentz-violation (see \cite{Kostelecky:2007kx}, but also \cite{Lehnert:2013jsa}).

The rationale for \cite{Kostelecky:2007kx} is that, if it were possible to find a background in which torsion can be taken as constant, with this approximation valid everywhere within the solar system, then the constant torsion background gives a preferred direction, and limits on Lorentz-violation convert into bounds on the presence of torsion terms.

These bounds are reliable only so long as the approximation is valid, and admittedly it is rather difficult to believe that the constant torsion background could stretch up to such large scales. A more physical situation would be to consider laboratory-sized experiments for the constant torsion background as those discussed in \cite{Lehnert:2013jsa}.

Then again, in this situation a full non-relativistic limit was taken, and it can be shown that this approximation kills the only component of torsion that was tested \cite{Fabbri:2014vda}.

Still, references \cite{Kostelecky:2007kx,Lehnert:2013jsa} are highly cited, and therefore a study of higher-order torsion interactions of spinor fields has to be done. In this paper, we do this by employing the VZ method of analysis \cite{Velo:1969bt,Velo:1970ur}. As in \cite{Fabbri:2014dxa}, propagations may be affected by non-causal behaviour in general.
\section{Interactions of Torsion with Derivatives of Spinors}
Throughout the work, we will consider the Dirac spinor indicated by the pair of conjugate $\overline{\psi}$ and $\psi$ fields, where the conjugation is $\overline{\psi}\!=\!\psi^{\dagger} \boldsymbol{\gamma}_{0}$ with $\boldsymbol{\gamma}_{0}$ the temporal Clifford matrix; with the full set of Clifford matrices $\boldsymbol{\gamma}^{a}$ we define also $\left[\boldsymbol{\gamma}^{a}\!, \!\boldsymbol{\gamma}^{b}\right]\!=\! 4\boldsymbol{\sigma}^{ab}$ and $2i\boldsymbol{\sigma}_{ab} \!=\!\varepsilon_{abcd}\boldsymbol{\pi}\boldsymbol{\sigma}^{cd}$ with $\boldsymbol{\pi}$ being the parity-odd Clifford matrix\footnote{The parity-odd matrix $\boldsymbol{\pi}$ is usually indicated as a gamma matrix with an index five, but this notation is misleading since the index five does not have any meaning in $4$-dimensions; and additionally, there may be a sign ambiguity according to whether the position of that index is upper or lower. Because the index five could lead to mistakes, without being really useful, we use a notation with no index at all. The choice of the letter is motivated by the fact that this matrix is connected to parity properties.}: these matrices verify the relation $\boldsymbol{\gamma}_{i}\boldsymbol{\gamma}_{j}\boldsymbol{\gamma}_{k}
\!=\!\boldsymbol{\gamma}_{i}\eta_{jk} \!-\!\boldsymbol{\gamma}_{j}\eta_{ik}
\!+\!\boldsymbol{\gamma}_{k}\eta_{ij}
\!+\!i\varepsilon_{ijkq}\boldsymbol{\pi}\boldsymbol{\gamma}^{q}$ which turns out to be an important identity we will use in the rest of the work. We will indicate with $\boldsymbol{\nabla}_{\mu}\psi$ the covariant derivative of spinors; this derivative will be taken in the torsionless case, because torsion can always be added as a supplementary field. Torsion is taken as the completely antisymmetric dual of an axial vector $W_{\sigma}$ as discussed in references \cite{A-L,M-L,Fabbri:2006xq,Fabbri:2009se,Fabbri:2014naa,Fabbri:2014dxa} and its covariant curl is indicated with $(\partial W)_{\alpha\nu}$ for the reason explained in reference \cite{Fabbri:2014dxa}.

We now have all the elements needed to construct the Lagrangian. Considering only parity-even terms gives
\begin{eqnarray}
\nonumber
&\mathscr{L}\!=\!
\frac{i}{2}(\overline{\psi}\boldsymbol{\gamma}^{\mu}\boldsymbol{\nabla}_{\mu}\psi
\!-\!\boldsymbol{\nabla}_{\mu}\overline{\psi}\boldsymbol{\gamma}^{\mu}\psi)
\!-\!\frac{1}{4}(\partial W)^{2}-\\
\nonumber
&-m\overline{\psi}\psi\!+\!\frac{1}{2}M^{2}W^{2}
\!-\!X\overline{\psi}\boldsymbol{\gamma}^{\mu}\boldsymbol{\pi}\psi W_{\mu}
\!-\!B\overline{\psi}\psi W^{2}+\\
\nonumber
&+Yi(\overline{\psi}\boldsymbol{\pi}\boldsymbol{\sigma}^{\mu\nu}\boldsymbol{\nabla}_{\mu}\psi\!-\!\boldsymbol{\nabla}_{\mu}\overline{\psi}\boldsymbol{\pi}\boldsymbol{\sigma}^{\mu\nu}\psi)W_{\nu}+\\
\nonumber
&+Y'\frac{1}{2}(\overline{\psi}\boldsymbol{\pi}\boldsymbol{\nabla}_{\mu}\psi\!-\!
\!\boldsymbol{\nabla}_{\mu}\overline{\psi}\boldsymbol{\pi}\psi)W^{\mu}+\\
&+A\overline{\psi}\boldsymbol{\pi}\boldsymbol{\sigma}^{\mu\nu}\psi(\partial W)_{\mu\nu}
\!+\!A'i\overline{\psi}\boldsymbol{\pi}\psi\nabla_{\mu}W^{\mu}
\label{l}
\end{eqnarray}
where $m$ and $M$ are the spinor and torsion masses while all others are the spin-torsion coupling constants, which are $X$ for the fourth-order term and $B$, $Y$, $Y'$, $A$, $A'$ for fifth-order terms; up to the fifth order this Lagrangian is the most complete, and higher-than-fifth-order terms are not included for no reasons other than simplicity.

Varying it with respect to the two fields involved gives
\begin{eqnarray}
\nonumber
&i\boldsymbol{\gamma}^{\mu}\boldsymbol{\nabla}_{\mu}\psi
\!-\!Y2iW_{\nu}\boldsymbol{\sigma}^{\nu\mu}\boldsymbol{\pi}\boldsymbol{\nabla}_{\mu}\psi
\!+\!Y'W^{\mu}\boldsymbol{\pi}\boldsymbol{\nabla}_{\mu}\psi+\\
\nonumber
&+(Y'/2\!+\!A'i)\nabla_{\mu}W^{\mu}\boldsymbol{\pi}\psi
\!+\!(iY/2\!+\!A)(\partial W)_{\mu\nu}\boldsymbol{\sigma}^{\mu\nu}\boldsymbol{\pi}\psi-\\
&-XW_{\mu}\boldsymbol{\gamma}^{\mu}\boldsymbol{\pi}\psi\!-\!BW^{2}\psi \!-\!m\psi\!=\!0
\label{sfe}
\end{eqnarray}
for the spinor and
\begin{eqnarray}
\nonumber
&\nabla_{\alpha}(\partial W)^{\alpha\nu}\!+\!(M^{2}\!-\!2B\overline{\psi}\psi)W^{\nu}
\!=\!X\overline{\psi}\boldsymbol{\gamma}^{\nu}\boldsymbol{\pi}\psi-\\
\nonumber
&-Yi(\overline{\psi}\boldsymbol{\pi}\boldsymbol{\sigma}^{\mu\nu}\boldsymbol{\nabla}_{\mu}\psi\!-\!\boldsymbol{\nabla}_{\mu}\overline{\psi}\boldsymbol{\pi}\boldsymbol{\sigma}^{\mu\nu}\psi)-\\
\nonumber
&-Y'\frac{1}{2}(\overline{\psi}\boldsymbol{\pi}\boldsymbol{\nabla}^{\nu}\psi\!-\!
\!\boldsymbol{\nabla}^{\nu}\overline{\psi}\boldsymbol{\pi}\psi)+\\
&+A\nabla_{\mu}(2\overline{\psi}\boldsymbol{\pi}\boldsymbol{\sigma}^{\mu\nu}\psi)
\!+\!A'\nabla^{\nu}(i\overline{\psi}\boldsymbol{\pi}\psi)
\label{tfe}
\end{eqnarray}
for the axial vector torsion. The last gives the constraint
\begin{eqnarray}
\nonumber
&(M^{2}\!-\!2B\overline{\psi}\psi)\nabla_{\nu}W^{\nu}
\!=\!2B\nabla_{\nu}(\overline{\psi}\psi)W^{\nu}+\\
\nonumber
&+X\nabla_{\nu}(\overline{\psi}\boldsymbol{\gamma}^{\nu}\boldsymbol{\pi}\psi)
\!+\!Y2i\boldsymbol{\nabla}_{\nu}\overline{\psi}\boldsymbol{\pi}\boldsymbol{\sigma}^{\nu\mu}
\boldsymbol{\nabla}_{\mu}\psi-\\
&-Y'\frac{1}{2}(\overline{\psi}\boldsymbol{\pi}\boldsymbol{\nabla}^{2}\psi\!-\!
\!\boldsymbol{\nabla}^{2}\overline{\psi}\boldsymbol{\pi}\psi)
\!+\!A'\nabla^{2}(i\overline{\psi}\boldsymbol{\pi}\psi)
\label{pcavc}
\end{eqnarray}
in the form of a partially-conserved axial-vector current.

These equations are next studied with the VZ method.
\section{Velo-Zwanziger Analysis}
As it is well known, the Velo-Zwanziger analysis \cite{Velo:1969bt,Velo:1970ur} is an instrument to check whether a given system of field equations displays inconsistency in the field propagation, such as lack of hyperbolicity and appearance of ellipticity or such as loss of causality and appearance of surfaces of wave-fronts that are boosted out of the light-cone.

To its full extent, the Velo-Zwanziger protocol goes as follows: 1. given a system of field equations, remove each term that does not contain the highest-order derivatives of the field; 2. in what remains, replace $i\boldsymbol{\nabla}_{\alpha}\!\rightarrow\!n_{\alpha}$ getting an equation in the form $\boldsymbol{A}\psi\!=\!0$ where $\boldsymbol{A}$ is an algebraic linear operator; 3. for this, require $\mathrm{det}\boldsymbol{A}\!=\!0$ in order to obtain an algebraic equation, called characteristic equation, which is discussed thusly: a. if in the characteristic equation the time component $n_{0}$ is allowed to have also complex solutions, then the original field equations cease to be hyperbolic and become elliptic; b. if the characteristic equation allows $n_{\alpha}$ to be time-like, then the original field equations cease to be causal \cite{Velo:1970ur}. As it is clear, field equations in the free case are always both hyperbolic and causal, and it is only when interactions are turned on that loss of hyperbolicity and causality might intervene, since interactions may give rise to the presence of extra terms containing the highest-order derivatives of the field.

The structure of the interaction that produces highest-order derivatives of the field may vary. But nevertheless, whenever the interaction term has the same order derivative of the kinetic term, problems are bound to happen.

This situation is the one we have here. The spinor field equations (\ref{sfe}) are precisely of this type, with interactions in $Y$ and $Y'$ displaying derivatives of the spinor field.

Using the VZ method on (\ref{sfe}), we get that the characteristic equation is given by the following determinant
\begin{eqnarray}
&\mathrm{det}(n_{\mu}\boldsymbol{\gamma}^{\mu}
\!-\!Y2W_{\nu}n_{\mu}\boldsymbol{\sigma}^{\nu\mu}\boldsymbol{\pi}
\!-\!Y'iW^{\mu}n_{\mu}\boldsymbol{\pi})\!=\!0
\end{eqnarray}
which has to be evaluated. A trick we can use is to have this determinant multiplied by $\mathrm{det}(\boldsymbol{\gamma}^{\alpha}n_{\alpha})\!\neq\!0$ which, being non-zero, would not change the information content of the above determinant, but it would simplify it to
\begin{eqnarray}
\nonumber
&\mathrm{det}[n^{2}\mathbb{I}\!+\!YW_{\nu}n^{2}\boldsymbol{\gamma}^{\nu}\boldsymbol{\pi}
\!-\!Y(W\!\cdot\!n)n_{\nu}\boldsymbol{\gamma}^{\nu}\boldsymbol{\pi}+\\
&+Y'(W\!\cdot\!n)n_{\nu}i\boldsymbol{\gamma}^{\nu}\boldsymbol{\pi}]\!=\!0
\end{eqnarray}
which can be computed; notice that according to whether we were to multiply by
$\mathrm{det}(\boldsymbol{\gamma}^{\alpha}n_{\alpha})$ on the left or right, the resulting determinant would be the same up to the sign of the $Y'$ term, but such a sign can always be re-absorbed with a complex conjugation, so that nothing will depend on it, as we will see. The straightforward computation of the above determinant gives the characteristic equation
\begin{eqnarray}
&n^{2}(1\!+\!Y^{2}W^{2})\!-\!(Y^{2}\!+\!Y'^{2})|W\!\cdot\!n|^{2}\!=\!0
\label{ce}
\end{eqnarray}
which is real and with no ambiguity of the $Y'$ sign, as we have promised. As clear, its solutions are always real but they may be time-like if $1\!+\!Y^{2}W^{2}\!>\!0$ which may always occur, provided that the torsion is weak enough.

Therefore, field equations (\ref{sfe}) are hyperbolic, but it is always possible to find situations in which they give rise to non-causal propagation. Causality is ensured, in every circumstance, when $Y\!=\!Y'\!=\!0$ thus suppressing higher-order interactions of torsion with derivatives of spinors. 

Moving to the study of the torsion field equations, it is possible to substitute the partially-conserved axial-vector current (\ref{pcavc}) into the field equation (\ref{tfe}), but in doing so we obtain a field equation in which there is no term that has the same order derivative of the kinetic term itself.

In this case therefore the VZ method would give that the leading term is the only term, which is thus the usual d'Alembert operator, yielding the characteristic equation in the form $n^{2}\!=\!0$ and henceforth showing that the initial field equations (\ref{tfe}) are always hyperbolic and causal.

This analysis says nothing about the other constants and hence the higher-order interactions of derivatives of torsion with spinors are still allowed in principle.
\section{Other Defects of Consistency}
We now examine the terms in the $A$ and $A'$ coefficients.

The spinor field equation is now reduced to
\begin{eqnarray}
\nonumber
&i\boldsymbol{\gamma}^{\mu}\boldsymbol{\nabla}_{\mu}\psi
\!+\!A'i\nabla_{\mu}W^{\mu}\boldsymbol{\pi}\psi
\!+\!A(\partial W)_{\mu\nu}\boldsymbol{\sigma}^{\mu\nu}\boldsymbol{\pi}\psi-\\
&-XW_{\mu}\boldsymbol{\gamma}^{\mu}\boldsymbol{\pi}\psi\!-\!BW^{2}\psi \!-\!m\psi\!=\!0
\end{eqnarray}
while the torsion field equation is 
\begin{eqnarray}
\nonumber
&\nabla_{\alpha}(\partial W)^{\alpha\nu}\!+\!(M^{2}\!-\!2B\overline{\psi}\psi)W^{\nu}
\!=\!X\overline{\psi}\boldsymbol{\gamma}^{\nu}\boldsymbol{\pi}\psi+\\
&+A\nabla_{\mu}(2\overline{\psi}\boldsymbol{\pi}\boldsymbol{\sigma}^{\mu\nu}\psi)
\!+\!A'\nabla^{\nu}(i\overline{\psi}\boldsymbol{\pi}\psi)
\end{eqnarray}
as it is easy to check. If the spinor field equation is multiplied on the left by $\boldsymbol{\pi}$ and contracted with the conjugate spinor, and the imaginary part is extracted, we get
\begin{eqnarray}
\nonumber
&\boldsymbol{\nabla}_{\mu}(\overline{\psi}\boldsymbol{\gamma}^{\mu}\boldsymbol{\pi}\psi)
\!=\!2(BW^{2}\!+\!m)i\overline{\psi}\boldsymbol{\pi}\psi+\\
&+2A'\nabla_{\mu}W^{\mu}\overline{\psi}\psi
\!-\!2A(\partial W)_{\mu\nu}i\overline{\psi}\boldsymbol{\sigma}^{\mu\nu}\psi
\end{eqnarray}
with partially-conserved axial-vector current
\begin{eqnarray}
\nonumber
&(M^{2}\!-\!2B\overline{\psi}\psi)\nabla_{\nu}W^{\nu}
\!=\!2B\nabla_{\nu}(\overline{\psi}\psi)W^{\nu}+\\
&+X\nabla_{\nu}(\overline{\psi}\boldsymbol{\gamma}^{\nu}\boldsymbol{\pi}\psi)
\!+\!A'\nabla^{2}(i\overline{\psi}\boldsymbol{\pi}\psi)
\end{eqnarray}
still being valid. These two can be combined to give
\begin{eqnarray}
\nonumber
&[M^{2}\!-\!2(B\!+\!XA')\overline{\psi}\psi]\nabla_{\nu}W^{\nu}
\!=\!2B\nabla_{\nu}(\overline{\psi}\psi)W^{\nu}+\\
\nonumber
&+A'\nabla^{2}(i\overline{\psi}\boldsymbol{\pi}\psi)
\!+\!2X(BW^{2}\!+\!m)i\overline{\psi}\boldsymbol{\pi}\psi-\\
&-2XA(\partial W)_{\mu\nu}i\overline{\psi}\boldsymbol{\sigma}^{\mu\nu}\psi
\end{eqnarray}
which can be substituted into the spinor field equations.

When this is done, the spinor field equations become
\begin{eqnarray}
\nonumber
&i\boldsymbol{\gamma}^{\mu}\boldsymbol{\nabla}_{\mu}\psi
\!+\!A'i[A'\nabla^{2}(i\overline{\psi}\boldsymbol{\pi}\psi)
\!+\!2B\nabla_{\nu}(\overline{\psi}\psi)W^{\nu}+\\
\nonumber
&+2X(BW^{2}\!+\!m)i\overline{\psi}\boldsymbol{\pi}\psi
\!-\!2XA(\partial W)_{\mu\nu}i\overline{\psi}\boldsymbol{\sigma}^{\mu\nu}\psi]\cdot\\
\nonumber
&\cdot[M^{2}\!-\!2(B\!+\!XA')\overline{\psi}\psi]^{-1}\boldsymbol{\pi}\psi
\!+\!A(\partial W)_{\mu\nu}\boldsymbol{\sigma}^{\mu\nu}\boldsymbol{\pi}\psi-\\
&-XW_{\mu}\boldsymbol{\gamma}^{\mu}\boldsymbol{\pi}\psi\!-\!BW^{2}\psi \!-\!m\psi\!=\!0
\end{eqnarray}
which have developed second-order derivatives of the field itself, and consequently they are unacceptable \cite{Velo:1970ur}.

Its removal necessitates that $A'\!=\!0$ hence leaving
\begin{eqnarray}
\nonumber
&i\boldsymbol{\gamma}^{\mu}\boldsymbol{\nabla}_{\mu}\psi
\!+\!A(\partial W)_{\mu\nu}\boldsymbol{\sigma}^{\mu\nu}\boldsymbol{\pi}\psi-\\
&-XW_{\mu}\boldsymbol{\gamma}^{\mu}\boldsymbol{\pi}\psi\!-\!BW^{2}\psi \!-\!m\psi\!=\!0
\end{eqnarray}
for the spinor field equations and  
\begin{eqnarray}
\nonumber
&\nabla_{\alpha}(\partial W)^{\alpha\nu}\!+\!(M^{2}\!-\!2B\overline{\psi}\psi)W^{\nu}
\!=\!X\overline{\psi}\boldsymbol{\gamma}^{\nu}\boldsymbol{\pi}\psi+\\
&+A\nabla_{\mu}(2\overline{\psi}\boldsymbol{\pi}\boldsymbol{\sigma}^{\mu\nu}\psi)
\end{eqnarray}
for the torsion field equations. There is no known manner with which to remove the $A$ term, as it is expected.

In fact, this term has the structure of a dipole moment and as such it is known to be perfectly allowed.

Albeit we can still have higher-order interactions of the derivatives of torsion with spinors, their effect identically vanishes whenever constant torsion is used \cite{Kostelecky:2007kx,Lehnert:2013jsa}.

Similarly the $B$ term gives no problem of consistency, but neither it serves any purpose for the Lorentz-violating effects because of its pure isotropic structure \cite{Lehnert:2013jsa}.
\section{Renormalizability Recovered}
Let us consider the above field equations after the reduction of the coefficients, and additionally let us remove the term in the $A$ coefficient so to focus on the $B$ term.

Straightforwardly, the spinor field equations are
\begin{eqnarray}
&i\boldsymbol{\gamma}^{\mu}\boldsymbol{\nabla}_{\mu}\psi
\!-\!XW_{\mu}\boldsymbol{\gamma}^{\mu}\boldsymbol{\pi}\psi\!-\!BW^{2}\psi \!-\!m\psi\!=\!0
\end{eqnarray}
and the torsion field equations are 
\begin{eqnarray}
&\nabla_{\alpha}(\partial W)^{\alpha\nu}\!+\!(M^{2}\!-\!2B\overline{\psi}\psi)W^{\nu}
\!=\!X\overline{\psi}\boldsymbol{\gamma}^{\nu}\boldsymbol{\pi}\psi
\end{eqnarray}
which are non-renormalizable, in principle.

In the effective approximation, the last becomes
\begin{eqnarray}
&(M^{2}\!-\!2B\overline{\psi}\psi)W^{\nu}
\!=\!X\overline{\psi}\boldsymbol{\gamma}^{\nu}\boldsymbol{\pi}\psi
\end{eqnarray}
which can be inverted, to have torsion substituted in the spinor field equation giving the final result
\begin{eqnarray}
\nonumber
&i\boldsymbol{\gamma}^{\mu}\boldsymbol{\nabla}_{\mu}\psi
\!+\!\frac{2X^{2}\phi^{2}}{(M^{2}\!-4B\phi^{2}\cos{\beta})}e^{i\beta\boldsymbol{\pi}}\psi+\\
&+\left[\frac{4BX^{2}\phi^{4}}{(M^{2}\!-4B\phi^{2}\cos{\beta})^{2}}\!-\!m\right]\psi\!=\!0
\end{eqnarray}
having used Fierz re-arrangements and writing the spinor as $\overline{\psi}\psi\!=\!2\phi^{2}\cos{\beta}$ and $i\overline{\psi}\boldsymbol{\pi}\psi\!=\!2\phi^{2}\sin{\beta}$ with scalar fields $\phi$ and $\beta$ known as module and Yvon-Takabayashi angle.

Of course, starting from non-renormalizable equations, and performing an effective approximation, one may well be led to assume that the final result would be even more dramatically non-renormalizable, but this is not at all the case here: as long as the Yvon-Takabayashi angle is not equal to $\pm\pi/2$ in general, large values of the module give
\begin{eqnarray}
&\!\!\!\!i\boldsymbol{\gamma}^{\mu}\boldsymbol{\nabla}_{\mu}\psi
\!-\!\left(\frac{X^{2}\tan{\beta}}{2B}\right)i\boldsymbol{\pi}\psi
\!-\!\left[\frac{X^{2}\cos{(2\beta)}}{4B(\cos{\beta})^{2}}\!+\!m\right]\psi\!=\!0
\end{eqnarray}
where the non-linear terms have become mass-like terms, which are known to have no problem of renormalization.
\section{Extensions}
In the construction of a Lagrangian, we need selection rules to limit the amount of interaction terms, and for a long time, it was believed that the single most important selection rule was renormalizability; despite there exists no mathematical way to justify this criterion, a heuristic argument can be made, and above all renormalizability seemed to work remarkably well. Nonetheless, some sign of fatigue has started to show, as for example in the case of Majorana mass terms, where the Yukawa-like coupling is a fifth-order dimensional term, but in this instance the lack of renormalizability has not been used to discourage its employment for neutrino mass generation mechanism.

In this paper we started from higher-order interactions, which are non-renormalizable, and we made an effective approximation, which gives effective interactions that are non-renormalizable, but we ended with interactions that were renormalizable: this is another example showing the renormalizability criterion as being too suppressive.

All throughout the paper we have considered only the axial vector dual of the completely antisymmetric part of torsion, but there are also other two parts of torsion that might be considered \cite{Capozziello:2001mq}: for instance, with the vectorial trace of torsion $T^{\mu}$ one may build $\overline{\psi}\psi\nabla_{\mu}T^{\mu}$ as a possible term in the interaction Lagrangian, and then apply this method to study the propagation. Because the effects of Lorentz violation were mostly considered to set limits on the completely antisymmetric part, and less on the other two parts, we are not going to deepen this extension, but it is clear that this extension is worth more investigation.

Finally, it is also worth mentioning that there are torsion theories in which torsion is not coupled to spins of spinor fields \cite{Cai:2015emx}, and experts in those domains might be willing to consider the present methods in order to study Lorentz violation in these alternative theories.
\section{CONCLUSION}
In this paper, we have analyzed all of the fifth-order dimensional interactions of torsion and spinor fields, and we have found that both terms containing derivatives of spinors violate causality; and the term that contains the divergence of torsion has analogous, thought somewhat worse, defects \cite{Velo:1970ur}. The term proportional to the curl of torsion is permitted; and the term algebraic in the square torsion is permitted too. We commented on the results.

In \cite{Kostelecky:2007kx,Lehnert:2013jsa} it has been claimed that specific types of interactions between torsion and spinor fields can produce, in specific contexts, a Lorentz-violation, so that limits on Lorentz-violation imply bounds on torsion: even within the validity of those specific contexts, these effects would be due to terms that are inconsistent. We discussed how the algebraic square-torsion term is non-renormalizable but still well-behaved at high energies, and how this is an example of physical situation for which the renormalizability criterion is much too constraining indeed.

We finally discussed possible extensions in terms of the two remaining parts of torsion, and alternative dynamics governing torsion coupling to material fields.

\end{document}